\documentclass[twocolumn,showpacs,preprintnumbers,amsmath,amssymb]{revtex4}

\usepackage{graphicx}  

\begin{document} 
\title{Spontaneous Symmetry Breaking in Two Coupled  Nanomechanical Electron Shuttles}
      
\author{Chulki Kim}

\author{Jonghoo Park}

\author{Robert H. Blick}
\email{blick@engr.wisc.edu}
\affiliation{
University of Wisconsin-Madison, 
Electrical \& Computer Engineering,
1415 Engineering Drive, Madison, 
WI 53706, USA.}

\date{\today}


\begin{abstract}
  {
  {\bf
  We present spontaneous symmetry breaking in a nanoscale version of a setup prolific in classical mechanics: two coupled nanomechanical pendulums. The two pendulums are electron shuttles fabricated as nanopillars~\cite{dvs_apl04}
and placed between two capacitor plates in a homogeneous electric field. 
Instead of being mechanically coupled through a spring they exchange electrons, i.e. they shuttle electrons from the source to the drain 'capacitor plate'. Nonzero DC current through this system by external AC excitation is caused via dynamical symmetry breaking. This symmetry-broken current appears at sub- and superharmonics of the fundamental mode of the coupled system.}}
\end{abstract}


\maketitle
   Spontaneous symmetry breaking is one of the fundamental ideas in elementary particle physics when unified gauge field theories are applied~\cite{weinberg}, e.g.
breaking the gauge symmetry leads to the difference in the electromagnetic and the weak interaction.  Commonly, one refers to a classical mechanics example to visualize  
symmetry breaking, such as a thin bar under pressure from both its clamping points.  Obviously, the equations of motion are invariant around the symmetry axis along the beam.
Once the force reaches a critical level the compressed bar will buckle and thus break the initial symmetry.  
Although, in the buckled state the bar will be in an energetic minimum,
the symmetry of this configuration is reduced.  From a quantum mechanical point of view the key question 
is into which of the many degenerate energetic minima the compressed bar will collapse to and what kind of fluctuation 
will have caused this transition.  Naturally, for a large bar governed by classical mechanics it would be hard to
associate the buckling with quantum mechanical states.  The situation changes drastically when one considers 
quantum electro-mechanical systems (QEM)~\cite{blencowe}, such as buckled 
nanomechanical beams which have been theoretically discussed by Carr~{\it et al.}~\cite{carr}.  In other words 
nanomechanical devices have the potential to address fundamental quantum mechanics~\cite{craighead,karrai,naik,steele}.

  In the following we will present electron transport measurements via two coupled electron shuttles at room temperature under AC/DC excitation. The two shuttles we fabricated have the same physical dimensions, are isolated from ground, and possess metallic tips.  
The actual realization of the device is shown in the scanning electron microscope graph in Fig.~1(a):  the two nanopillars forming the shuttles are placed in 
series between source and drain contacts.  The diameters of the islands on top of each nanopillar is 65~nm with pillar heights of 250~nm.
The inter-pillar distance is 17~nm and is much less than the gaps to source and drain, enhancing mode coupling.
The distance from source to the first pillar and drain to the second one are identical, making the setup indeed very symmetric. 
Our particular focus will be on how breaking the local symmetry leads to the rectified current as recently
conjectured by Ahn~{\it et al.}~\cite{ahn}. 
The fundamental outcome is that although both the nanomechanical and the nanoelectronic circuits are operated in the linear regimes an AC voltage can be rectified. Commonly, rectification only occurs based on a nonlinearity, however, such symmetry breaking in coupled nanomechanical pendulums is exactly what is predicted in Ref.~\cite{ahn}.

\begin{figure}[!htbp]
\includegraphics[width=3.4in]{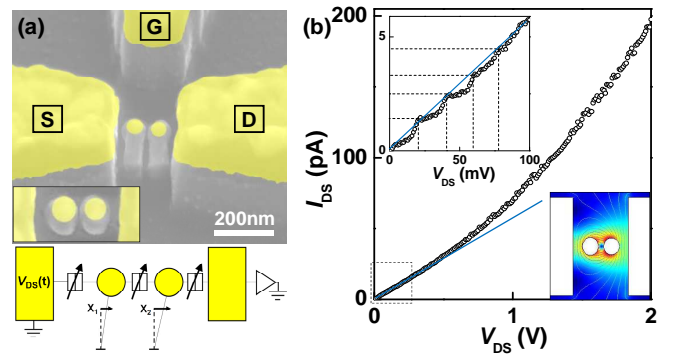}
\caption{(a) Aerial view in a scanning electron microscope of two coupled electron shuttles.
The two nanopillars' gold tops have diameters of $\cong$~65~nm with a total nanopillar height of around 250~nm.
The inset gives a top-view of the coupled nano-electromechanical pendulum. The scale bars are 200~nm in both graphs.
Below: sketch of the measurement  setup:  an applied AC and DC bias voltage ($V_{\rm DS}$) displaces the two pendulums by $x_1, x_2$ and the resulting current $I_{\rm DS}$ is amplified.  
(b) Direct current $I_{\rm DS}$ vs. bias voltage $V_{\rm DS}$ in the case of pure self-excitation, 
i.e. $V_{\rm AC} = 0$ at room temperature. The upper inset shows low bias regime with the dashed lines indicating 
the occurrence of Coulomb charging effects (see text for details).  The lower inset gives a finite-element simulation of the
electric field distribution under an applied bias voltage (top view). The field intensity is maximal between the two pillars.} 
 \label{fig1}
\end{figure}
  The two nanopillars are milled out of a silicon-on-insulator (SOI) material,
where the top crystalline silicon is 190~nm thin and the insulating SiO$_2$ is about 350~nm thick. A 50~nm top gold layer serves as the electrical conduction path. The deposited metal serves as an etch mask in a dry etch step which mills out the SOI material around the pillars. We apply a CF$_4$ plasma etch step and etch into the SiO$_2$ insulating layer, thus ensuring electron transport via the metallic islands (further processing details are given in Ref.~\cite{hskim07}).  The nanoscale circuit is placed in an impedance matched transmission line in order to minimize signal loss along the line.   
All measurements are performed under vacuum in a probe station at room temperature.  The station is placed 
in a Faraday cage and is equipped with radio frequency contact probes covering the range from DC to 50 GHz ( a bias-tee allows AC/DC superposition 
with high precision). The equivalent circuit diagram is given in the lower part of Fig.~1(a):  the two pillars can be individually displaced by $x_{1}$ and
$x_{2}$, which lead to tunable resistances and capacitances (arrow boxes). 

  In the first measurement we trace the direct current at different voltage bias values at room temperature as shown in Fig.~1(b).
At first glance the $IV$-characteristic appears to be linear with only a small deviation ($\propto V^2$) at above 1~V. 
Electron shuttling is leading to the non-zero current, which is caused by self-excitation (SE) as we have shown earlier
for single nanopillars~\cite{hskim09}. This can also be shown for lateral electron shuttles~\cite{dvs_apl04,koenig}. 
In this particular case we assume that several mechanical modes of the coupled
nanopillars (see below in Fig. 2) support the current. The ease of excitation is supported by the finite-element numerical simulation
shown in the inset of Fig.~1(b). Obviously, the electric field intensity is dramatically enhanced around the two pillars, enabling the onset
of mechanical oscillations under an applied DC bias voltage.
Closer inspection of the low-voltage regime reveals the onset of 
Coulomb blockade (CB) in these serially coupled electron shuttles, as predicted by Gorelik~{\it et al.}~\cite{gorelik98}.
The dashed lines in the inset indicate the individual Coulomb staircase, which is getting weaker towards voltages above 100~mV.
A brief investigation of the trace reveals
that the CB charging energy is of the order of the room temperature energy at~26~meV. 
Since we are operating the coupled shuttles at large AC-voltage swings ($> 100$~mV amplitude) as compared to the thermal background 
($V_{\rm th} \cong 26$~mV), we can focus on the overall linear $IV$-characteristic. For all measurements presented here the gate voltage potential was left at ground. 

\begin{figure}[!htbp]
\includegraphics[width=3.4in]{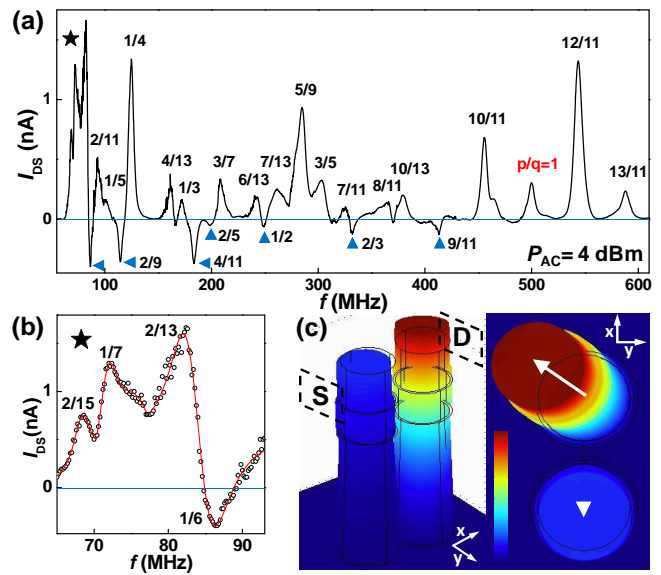}
\caption{(a) Spectrum of drain current $I_{\rm DS}$ vs. $f$ the excitation frequency of the coupled nanopillars. 
The current resonances are labeled according to the commensurate mode numbers $p/q$ where the fundamental mode is
found to be at $\omega_o/(2 \pi) = 504$~MHz from finite element simulations.  
The externally applied power of the radio frequency source was $P = 4$~dBm. Note the seven distinct current peaks 
in the reverse current (blue triangles). This pumping electro-mechanical pumping mechanism pushes electrons against the DC
bias of $V_{\rm DS} = 0$~mV. The low-frequency region ($\star$) is plotted in detail in (b). The red dashed lines are fits to the resonances
taking into account different line shapes. (c) Finite element simulation based on COMSOL showing the fundamental mode of the coupled pendulums in a 3D (left) and top view (right).  The color coding indicates the displacement within the nanopillars.} 
\label{fig2}
\end{figure}
  A radio frequency signal in the range of 1~MHz to 1~GHz is applied via
 high-frequency probes in order to trace the mechanical spectrum of the coupled nanopillars, similar to earlier work on single nanopillars~\cite{dvs_apl04,hskim07}.  The resulting output current from the coupled electron shuttles is a time-averaged DC signal leading
 to the spectrum shown in Fig.~2. The DC response of the coupled shuttles to the AC excitation is plotted at a DC bias voltage of $V_{\rm DS} = 0$~mV.
A broad set of resonances can be identified representing different modes of the coupled nanopillars. On closer inspection of the overall spectrum we find that the multitude of peaks arises due to a parametric instability caused 
by dynamical symmetry breaking, as predicted by Ahn~{\it et al.}~\cite{ahn}. The ratios noted at each peak follow $(p/q)\omega_0/(2 \pi)$ with $p,q$ being integers. From finite element simulations we find that the most likely fundamental mode of the coupled nanopillars is at around $\omega_o/(2 \pi) = 504$~MHz at $p/q = 1$, see Fig.~2(a). Unlike the theoretical prediction~\cite{ahn}, we find both even and odd numbers for $q$. The hierarchy of resonances is also slightly different from the theoretical result in that the current spectrum of the coupled electron shuttles has a richer structure with higher orders for $p$. We can attribute these differencies to more degrees of freedom in the real device allowing mechanical modes in all three spatial dimensions. 

 With increasing external AC power we observe peak broadening, overlaping with neighboring peaks. We find the modes with the winding numbers, $p/q$ generated by 'Farey sums' of neighboring modes' fractions when overlapping on the left and right: The numerator is the sum of the numerators of the fractions on either side, and the denominator is the sum of the denominators of the fractions on either side~\cite{farey1}. For instance, $p/q=1/7=(2+2)/(15+13), 2/13=(1+1)/(7+6), 2/11=(1+1)/(6+5)$ and so on. A similar behavior was observed in nonlinear systems showing structure of Arnold tongues~\cite{farey2}. The appearance of fractions of the fundamental mode and Farey trees with neighboring modes' fractions indicate that we detect dynamical symmetry breaking. 

 Most peaks of the spectrum show a forward current, as one expects under forward bias, while some of the peaks give a reverse current(blue triangles). This remarkable feature of the coupled shuttles, i.e. reverse shuttling or pumping, marks a bistability depending on initial conditions, such as positions and velocities of two islands at a given time. We need to note, as will be underlined theoretically below, that the resulting rectified current at zero bias observed here is solely due to a dynamical symmetry breaking and not by a mechanical or electronic nonlinearity. In detail the resonance shapes of the two-nanopillar system are analyzed in Fig.~2(b): a magnified part of the spectrum at below 100~MHz is shown with three fits (red lines) indicating the different line shapes. The resonance around 69~MHz follows the conventional Lorentzian, while the trace at 84~MHz shows a maximum and minimum.

\begin{figure}[!htbp]
\includegraphics[width=3.4in]{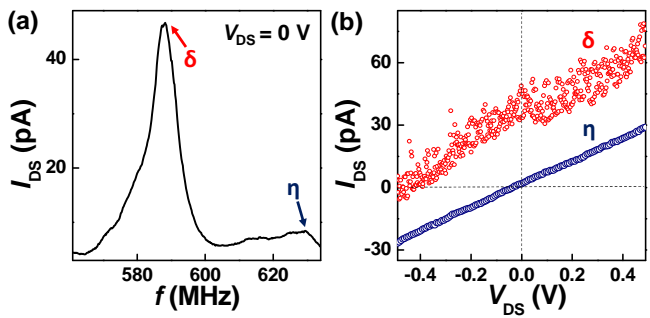}
\caption{(a)  Current response on- and off-resonance:
at 589~MHz ($\delta$) with an AC radio frequency excitation of -3~dBm a strong mechanical response of the coupled
pendulums is obtained, while at off-resonance ($\eta$) only a weak current shows. 
(b) Bias dependence of the coupled nanopillars on-resonance $\delta$ and off-resonance $\eta$.  While we find an ohmic
response for both cases the red $\delta$-trace is offset from the blue $\eta$-trace.  This offset results from the rectification through 
the coupled nanomechanical shuttles. } \label{fig3}
\end{figure}

 We now turn to the details of the rectification effect due to symmetry breaking: as an example we select the pronounced resonance at 
 $(13/11)\times \omega_0$. The particular peak around 589~MHz is plotted in Fig.~3(a) and marked as $\delta$. The off-resonance background is labeled with $\eta$. In Fig.~3(b) the $IV$-traces for these two frequency points are given: as expected for the background the $\eta$-line shows a linear dependence 
crossing through the origin of the graph. In contrast the $\delta$-trace reveals a clear offset under finite AC excitation at 589~MHz. 
This motion of the coupled charge shuttle indicates the existence of a bifurcation point~\cite{ahn}. 

  The theoretical underpinning for this bifurcation of the coupled shuttle leads to the rectified current~\cite{ahn}. 
A sinusoidal electromagnetic excitation $V_0 \sin(\omega t)$ is applied to the source contact, inducing mechanical displacement of   the shuttle at their eigenfrequencies. 
We can assume that the capacitances between the electrodes are constant, while the tunneling resistance $R$ varies 
 with electrode distance as 
$R_1(x_1)=R_1(0)e^{x_1/{\lambda}},  R_2(x_1-x_2)=R_2(0)e^{(x_2-x_1)/{\lambda}},  R_3(x_2)=R_3(0)e^{-x_2/{\lambda}}$, where $\lambda$ is the tunneling length. The $x_j$ refer to the three junctions ($j = 1,2,3$) in series, 
 i.e. from source to the first nanopillar, in between the pillars, and finally from the second pillar towards drain. 
The coupled nanopillars can be assumed to show a {\it linear} response under modest excitation as measured in Fig.~1(b) leading to a mechanical displacement, which can be described by a differential equation in the adiabatic limit~\cite{ahn}
 
 \begin{equation}
\ddot{x} - \gamma \dot{x} - \omega^2_{0} x = - \frac{c V_0^2 \sin \omega t}{m L} \tanh \left[ \frac{3x}{4 \lambda}\right], 
\end{equation}

 where $x=x_1-x_2$ is the relative coordinate, $\gamma$, $\omega_0$ denote the friction coefficient and the natural angular frequency of the shuttles, $m$ gives the nanopillars mass, and $L$ is the source-drain distance. There is a pair of bistable solutions, $x(t)$ and $-x(t)$. The time-averaged direct current is found to be 

\begin{equation}
I_{\rm dc} = \frac{\omega V_0}{4 \pi R} \int_{t_0}^{t_0 + 2 \pi/ \omega} \frac{\sin \omega t}{e^{x(t)/2\lambda} + e^{-x(t)/2\lambda}} dt,
\end{equation}

where we assume a symmetric configuration: $R_1(0)=0.5R_2(0)=R_3(0)\equiv R$. For a single nanopillar with a symmetric source-drain the direct current with AC input voltage will always average out to zero, making the above integral zero. However, for two coupled nanopillars the symmetry will be broken leading to a DC signal as we observe in Fig.~3(b). We note that this current only can be found for two coupled nanopillars.  
 
\begin{figure}[!htbp]
\includegraphics[width=3.4in]{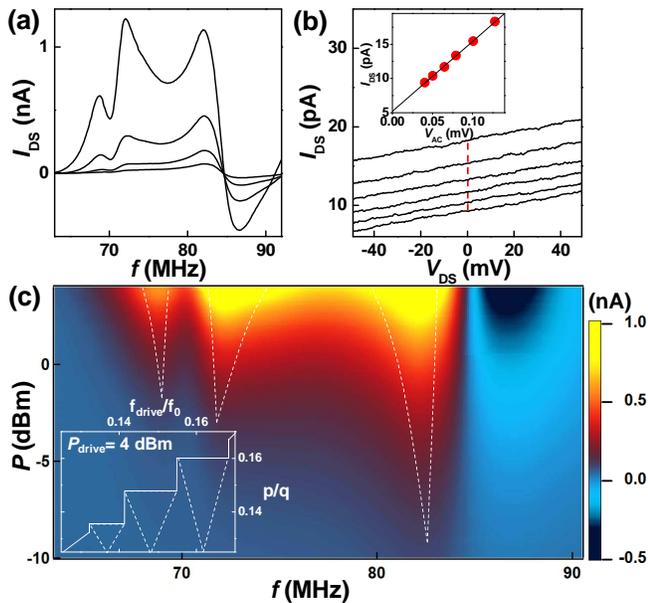}
\caption{Arnold tongues:  (a) mechanical resonances at around 80~MHz under increased AC power. 
(b) The DC bias dependence with gradually increased AC power, from -20~dBm (bottom) to -15~dBm (top). Inset shows the linear dependence of the rectified direct current upon increasing AC power.
(c)  Color scale plot of the total power dependence, revealing Arnold's tongues with the whited dashed lines being guides to the eye. Inset shows Devil's staircase plot depicting the width of each bistable region.
} 
\label{fig4}
\end{figure}
  In order to further verify the dynamical symmetry breaking and the resulting rectification we determined the AC power dependence of the low-frequency end at around 80~MHz, revealing the expected Arnold tongues~\cite{arnold}. Fig.~4(a) shows the detailed power dependence of the output current with line plots with the power level increasing from -10~dBm to +4~dBm. 
In Fig.~4(b) the maximal current at 83~MHz is plotted vs. DC bias and AC power ramped up as the parameter. The inset gives the scaling of generated direct current with respect to applied AC-power, indicating the rectification.
In Fig.~4(c) finally the total power dependence is shown in a color scale plot: 
the boundaries of each tongue are seen. The experimental bifurcation diagram shows the gradual appearance of higher order resonances as the input power increases. The thin dashed lines trace the increase in width as expected for Arnold's tongues. The appearance of Arnold's tongues is also observed in mathmatical simulations of two charge shuttles coupled via electron tunneling as discussed in Ref.~\cite{ahn}. In each bistable region the fundamental mode and the external excitation are mode-locked in spite of the frequency mismatch. Devil's staircase depicts winding numbers of the mode-locked regions, as shown in the inset of Fig.~4(c). This staircase maps the widths of the Arnold's tongues, defined as the FWHM for a given power of AC excitation. Even though this two pillar system does not have a perfect left-right symmetry, driving the enhanced net current was possible due to the robustness of the observed mode locking.    

  In summary we find dynamical symmetry breaking in coupled linear nanopillars traced in the current response of the pillars to an AC excitation. The identified resonances were agreed with the theoretical expectation by~\cite{ahn} except for some discrepanies due to the three-dimensional nature unlike the calculation model. The power dependance manifests the Arnold tongue structure where the bistability of the system allows for nonzero currents. This setup has potential as a bifurcation amplifier for signal processing applications~\cite{bif_amp} and can prove to be useful for energy scavenging applications~\cite{rectenna}. 
  
\vspace{1cm}

{\bf Acknowledgements --} 
The authors like to thank DARPA for support through the NEMS-CMOS program (N66001-07-1-2046) and the Graduate School of the University of Wisconsin-Madison for support through a Draper-TIF award.



\end{document}